\documentclass[aps,prl,twocolumn]{revtex4}%
\usepackage{amsmath}
\usepackage{graphicx}
\usepackage{dcolumn}
\usepackage{bm}%
\usepackage{amsfonts}%
\usepackage{amssymb}
\usepackage{color}
\begin{document}
\title{Valence Force Model for Phonons in Graphene and Carbon Nanotubes}
\author{Vasili Perebeinos} \email{vperebe@us.ibm.com}
\affiliation{IBM Research Division, T. J. Watson Research Center,
Yorktown Heights, New York 10598}
\author{J. Tersoff} \email{tersoff@us.ibm.com}
\affiliation{IBM Research Division, T. J. Watson Research Center,
Yorktown Heights, New York 10598}
\date{\today}

\begin{abstract}
Many calculations require a simple classical model for the
interactions between $sp^2$-bonded carbon atoms, as in graphene or
carbon nanotubes. Here we present a new valence force model to
describe these interactions. The calculated phonon spectrum of
graphene and the nanotube breathing-mode energy agree well with
experimental measurements and with \textit{ab initio} calculations.
The model does not assume an underlying lattice, so it can also be
directly applied to distorted structures. The characteristics and
limitations of the model are discussed.

\end{abstract}

\maketitle

Graphene and carbon nanotubes are remarkable materials, notable for
both their fascinating properties and their technological promise
\cite{Avouris}. In both contexts, it is often necessary to calculate
the phonons for problems where the use of \textit{ab initio} methods
is not feasible. For graphitic systems, this has usually been
approached by approximating the force-constant matrix with terms
coupling atoms up to some maximum distance
\cite{Jishi1,Jishi2,Lu,Saito1,Mohr1,Dubay,Dobarzik}. This approach
has many appealing features, but it has two important limitations.
First, the terms in the force-constant matrix decay smoothly with
distance between atoms \cite{Harrison}, so in practice it is
necessary to truncate the expansion long before it has converged.
Second, this approach is generally restricted to describing phonons
in the ideal crystal. It has required some ingenuity and
inconvenience even to extend these models to nanotubes, based on an
idealized curved-graphene structure
\cite{Jishi2,Saito1,Lu,Dobarzik}.

In order to treat phonons in large low-symmetry systems, such as
rumpled graphene or bent nanotubes, one would like a model that
explicitly gives the energy as a function of atomic positions,
without reference to any underlying crystal structure.
In principle one could use
the general-purpose empirical interatomic potentials that are
available for carbon, such as Ref.~\cite{Tersoff}.  But phonon
applications typically require higher accuracy than such
general-purpose models can provide.

For diamond- and zincblende-structure semiconductors, the problem
was largely solved by the use of ``valence force'' models. These
models use a smaller number of more complex terms, which may be more
or less physically motivated \cite{Kane}. However, to date only one
valence force model has been proposed for graphene
\cite{Aizawa,Rubio}; and it explicitly references the graphene
lattice, hindering application to distorted structures
\cite{Popov1}.

Here we present a new valence force model for $sp^2$-bonded carbon.
The model explicitly gives energy as a function of atomic positions,
without reference to any underlying crystal structure. The only
restriction is that the local geometry be consistent with $sp^2$
bonding, i.e.\ three neighbors not too far from 120$^\circ$ degrees
apart. Thus it can be directly applied not only to graphene, but to
nanotubes and fullerenes, in relaxed or distorted configurations. We
have tested the model for phonons in graphene and carbon nanotubes.
We first describe the model itself, and the fitting procedure. We
then present and discuss the phonon spectrum which is obtained after
fitting the model parameters to selected experimental and
theoretical data. Finally we discuss the overall accuracy and
limitations, along with some related issues such as anharmonicity.

We write the energy as%
\begin{align}
E &  =\beta_{r1}r_{0}^{-2}\sum_{i,j\in i}\left(  \delta r_{ij}\right)
^{2}+\beta_{c}\sum_{i,j<k\in i}\left(  \delta c_{i,jk}\right)  ^{2}
\nonumber \\
& +\beta_{v}r_{0}^{-2}\sum_{i,j<k<l\in i}\left(  \frac{3v_{ij}\cdot
v_{ik}\times v_{il}}{r_{ij}r_{ik}+r_{ik}r_{il}+r_{il}r_{ij}}\right)  ^{2}
\nonumber\\
&  +\beta_{r2}r_{0}^{-2}\sum_{i,j<k\in i}\left(  \delta r_{ij}\right)
\left(  \delta r_{ik}\right)
+\beta_{p}\sum_{i,j\in i}\left\vert \pi_{i}
\times \pi_{j}\right\vert ^{2}  \nonumber \\
& +\beta_{rc}r_{0}^{-1}\sum_{i,j\ne k < l\in i}\left(  \delta r_{ij}\right)
\left( \delta c_{i,kl}\right)
\label{energy}%
\end{align}
where $v_{ij} =v_{j}-v_{i}$, $v_{i}$ being the atomic position vector
of atom $i$, and the bondlength is $r_{ij}=\left\vert v_{ij}\right\vert $.
In the summations, $j \in i$ means $j$ runs over three neighbors of atom $i$,
$j<k \in i$ means  $j$ and $k$ are both neighbors of $i$ (ordered to avoid double counting),
restriction $j<k<l$ leaves only one possibility for the three neighbors of $i$, while restriction
$j\ne k<l$ gives three terms for each $i$.

The bond length in the ground state of graphene is $r_{0}$=0.142 nm;
$\delta r_{ij}=r_{ij}-r_{0}$. We further define
\begin{align}
\delta c_{i,jk} & = \frac{1}{2} + \frac{ v_{ij}\cdot v_{ik} }{r_{ij}r_{ik} }
\nonumber \\
\pi_{i} & =3 \frac{  v_{ij}\times v_{ik}+v_{ik}\times
v_{il}+v_{il}\times v_{ij}}{r_{ij}r_{ik}
+r_{ik}r_{il}+r_{il}r_{ij}}
\label{param}%
\end{align}
where $j$, $k$, and $l$ are the three neighbors of $i$.

The first two terms in Eq.~(\protect{\ref{energy}}) represent the
bond stretching stiffness $\beta_{r1}$ and bending stiffness
$\beta_{c}$, as in the Keating model \cite{Keating}. However, the
form here avoids the large and unphysical anharmonicities of the
Keating model. The third term $\beta_{v}$ provides stiffness against
out-of-plain vibrations. The fourth term $\beta_{rc}$ is motivated
by bond-order potentials \cite{Tersoff}. The fifth term $\beta_{p}$
gives stiffness against misalignment of neighboring $\pi$ orbital.
The last term $\beta_{rc}$ couples bond stretching and bond bending.

\begin{figure}[hb]
\includegraphics[width=3.5in]{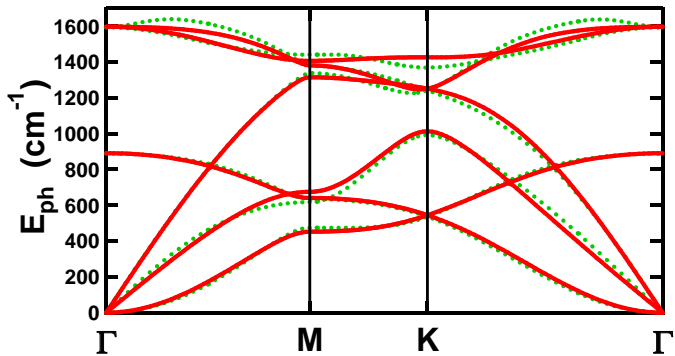}
\caption{Phonon band dispersions.
Green dotted curves are LDA calculations \protect{\cite{Dubay}}.
Red solid curves are results of the model Eq.~(\protect{\ref{energy}})
fitted to these LDA calculations.
The corresponding parameters are given in \protect{\cite{footnote1}}.}
\label{fig-fit1}%
\end{figure}

In fitting such a model, one typical chooses a set of data that it
is desired to reproduce, and defines a weighted error which is to be
minimized. As a straightforward test of the model and its ability to
reproduce realistic phonon dispersions, we first try fitting to
published LDA calculations \cite{Dubay,footnote1}. The result is
shown in Fig.~\ref{fig-fit1}. [We follow the spectroscopic
convention of reporting phonon energies in cm$^{-1}$, where 1
cm$^{-1}$ means $hc$/(1\,cm) $\approx$ 0.124 meV.] The rms error is
only 22.6 cm$^{-1}$, substantially less than the best previous fit
to GGA dispersions using a valence force model with five parameters
\cite{Rubio}.

By giving more weight to one feature or another in the fitting, it
is straightforward to improve the description of e.g.\ the acoustic
branches at the cost of worsening the optical branches. However,
regardless of how we weighted the data, we could not reproduce the
dips in the highest phonon bands at $\Gamma$ and $K$ while keeping a
reasonable overall dispersion. This issue was also mentioned in
\cite{Rubio}. Electron-phonon interactions are known to affect
phonon dispersions even in bulk semiconductors \cite{Harrison}; and
such interactions are particularly important for the highest bands
of graphene near $\Gamma$ and $K$ due to the Kohn anomaly
\cite{Ferrari}. Thus we cannot expect to describe these dips with
short-range classical interactions. It would therefore seem logical
to fit the bands away from $\Gamma$ and $K$, and accept that the
model gives energies that are too high for the top bands at those
points.  However, because the optical phonon energy at $\Gamma$ is a
widely used reference, we have chosen to fit this point accurately.

\begin{figure}[hb]
\includegraphics[width=3.5in]{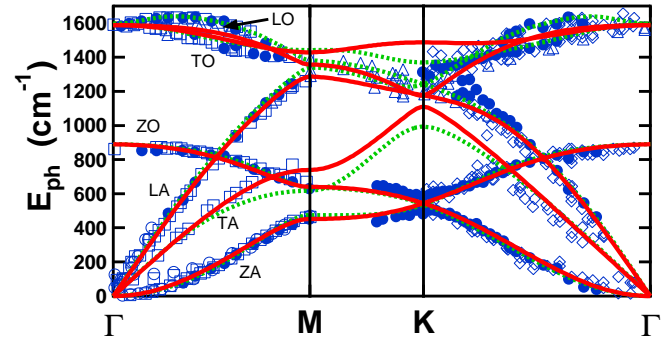}
\caption{Phonon band dispersion, comparing fitted model
with experimental data and \textit{ab initio} calculations.
Red solid curve is our model, Eq.~(\protect{\ref{energy}}),
with the parameters given in Table \protect{\ref{tab1}}.
Green dotted curve is an LDA calculation \protect{\cite{Dubay}}.
Blue symbols are experimental data:
electron energy loss spectroscopy (EELS) from Refs.~\protect{\cite{Oshima}},
\protect{\cite{Siebentritt}}, and  \protect{\cite{Yanagisawa}} (respectively
squares, diamonds, and filled circles), neutron scattering from
Ref.~\protect{\cite{Nicklow}} (open circles), and x-ray scattering from
Ref.~\protect{\cite{Maultzsch}} (triangles).  Data for
Refs.~\protect{\cite{Oshima,Siebentritt,Yanagisawa,Nicklow,Maultzsch}}
are taken from Ref.~\protect{\cite{Marzari}}.
}
\label{fig-fit}%
\end{figure}

We find that the Poisson ratio calculated with our model fitted to
the LDA calculations alone is $\nu=0.4$, much less than the
experimental value of $\nu=0.17$. This suggests that elastic
properties should be included in the fitting.  Also, the
experimental and theoretical data are not in perfect accord. We have
therefore chosen to fit a mixture of published experimental phonon
data, \textit{ab initio} phonon calculations, and elastic constants.
The resulting parameter values are listed in Table~\ref{tab1}, and
the corresponding phonon dispersion is shown as a solid curve in
Figure~\ref{fig-fit}. The corresponding elastic constants are given
in Table~\ref{tab2}. (We equate in-plane elastic properties of
graphene and graphite using the experimental layer spacing $c=6.7$
\AA \ and volume per atom $V_0=3\sqrt{3}r_0^2c/8$.)

Overall we consider the agreement in Figure~\ref{fig-fit}
and Table~\ref{tab2} to be quite good. The quality
of the fit is a highly nonlinear function of the parameters, so
there may be entirely different sets of parameters that give a
similar or even better agreement with the same data.

\begin{table}[hb]
\caption{\label{tab1} Parameters of the model Eq.~(\protect{\ref{energy}})
used in Fig.~\ref{fig-fit}, based on best fit to the experimental data
and LDA calculations.  Units are eV.}
\begin{ruledtabular}
\begin{tabular}{cccccc}
 $\beta_{r1}$& $\beta_{c}$ & $\beta_{v}$ & $\beta_{r2}$ & $\beta_{p}$ & $\beta_{rc}$ \\
\hline
18.52 & 4.087 & 1.313 & 4.004 & 0.008051 & 4.581 \\
\end{tabular}
\end{ruledtabular}
\end{table}

The longitudinal and transverse sound velocities ($v=d\omega/dq$ at
$q=0$) within our model are
\begin{eqnarray}
M_Cv_{T}^2&=& \frac{81}{4}\frac{4\beta_{r1}\beta_c-2\beta_{r2}\beta_{c}-\beta_{rc}^2}
{12\beta_{r1}+27\beta_c-6\beta_{r2}-18\beta_{rc}}
\nonumber \\
M_Cv_{L}^2&=&M_Cv_{T}^2+\frac{3}{2}\left( \beta_{r1}+\beta_{r2}\right)
\label{veloc}
\end{eqnarray}
where $M_C$ is the mass of a carbon atom. The model velocities
$v_{TA}\approx 13.3$ km/s and $v_{LA} \approx 21.2$ km/s are very
close to the experimental values of $v_{TA}\approx 14$ km/s and
$v_{LA}\approx 24$ km/s \cite{Oshima}. The elastic constants are
related to the sound velocities:
$V_0C_{66}=M_Cv_T^2$ and $V_0C_{11}=M_Cv_L^2$ \cite{Landau}.
\begin{table}[hb]
\caption{\label{tab2} Elastic constants from the experiment
 and the model (in GPa). Note a relation among elastic constants
\cite{Landau,Popov1} for hexagonal symmetry:
$C_{66}=(C_{11}-C_{12})/2$, $\nu=C_{12}/C_{11}$. }
\begin{ruledtabular}
\begin{tabular}{cccccc}
 & C$_{11}$& C$_{12}$ & C$_{66}$ & $\nu$ \\
\hline
experiment \cite{elast1} & 1060 &  180 &  440 &  0.17 \\
model  & 1024 &  210 &  407 &  0.20 \\
\end{tabular}
\end{ruledtabular}
\end{table}

\begin{figure}[hb]
\includegraphics[width=3.5in]{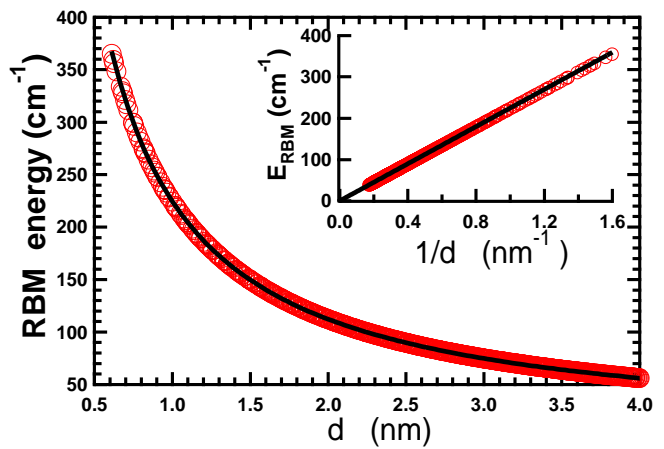}
\caption{Radial Breathing Mode (RBM) energy as a function of tube diameter (red open circles) along with the best fit
$\hbar\omega_{RBM}=224.6$ cm$^{-1}/d$ (black solid curve). The inset shows the same results versus inverse diameter.}%
\label{fig-rbm}%
\end{figure}

It is often convenient to have analytic expressions for the phonon
energies at symmetry points (e.g.\  for verifying a numerical
implementation). From Eq.~(\protect{\ref{energy}}),
\begin{eqnarray}
\omega^2&=& \left(M_Cr_0^2 \right)^{-1}\sum_{i}\alpha_{i}\beta_{i}
\label{eqgen}
\end{eqnarray}
where the index $i$ runs over $(r1,c,v,r2,p,rc)$ and coefficients
$\alpha_i$ are given in Table~\ref{tab3}.

\begin{table}[hb]
\caption{\label{tab3} Exact coefficients $\alpha_i$ for the
analytical expressions Eq.~(\protect{\ref{eqgen}}) of phonon energies
at high symmetry $k$ points.}
\begin{ruledtabular}
\begin{tabular}{ccccccccc}
$k$ point  & mode & $\hbar\omega$ cm$^{-1}$ & $\alpha_{r1}$ & $\alpha_{c}$ & $\alpha_{v}$ &
$\alpha_{r2}$ & $\alpha_{p}$ & $\alpha_{rc}$ \\
\hline
$\Gamma$ & ZO &  889  & 0  & 0 & 54 & 0 & 0 & 0  \\
 & LO/TO &  1588  & 12  & 27 & 0 & -6 & 0 & -18  \\
\hline
M & ZA &  452  & 0  & 0 & 6 & 0 & 1296 & 0  \\
 & ZO &  640  & 0  & 0 & 24 & 0 & 648 & 0  \\
 & TA &  740  & 0  & 12 & 0 & 0 & 0 & 0  \\
 & LA &  1286  & 8  & 0 & 0 & 0 & 0 & 0  \\
 & LO &  1357 & 4  & 27 & 0 & 2 & 0 & -6  \\
 & TO &  1429  & 12  & 3 & 0 & -6 & 0 & -6  \\
\hline
K & ZO/ZA &  544  & 0  & 0 & 13.5 & 0 & 1093.5 & 0  \\
 & TA &  1110  & 0  & 27 & 0 & 0 & 0 & 0  \\
 & LO/LA &  1177  & 6  & 6.75 & 0 & 1.5 & 0 & -4.5  \\
 & TO &  1487  & 12  & 0 & 0 & -6 & 0 & 0  \\
\end{tabular}
\end{ruledtabular}
\end{table}

Turning from graphene to carbon nanotubes, we calculate the radial
breathing mode  (RBM) for tubes of different diameter and chirality.
This mode corresponds to a radial stretching or compression of the
tube.  The mode emerges from the lowest-energy acoustic phonon modes
in graphene. The RBM acquires finite energy at zero wavevector due
to the nanotube curvature, with a simple $\approx 1/d$ scaling of
energy with diameter.  As a result, RBM measurements are widely used
to identify the diameters of single-walled carbon nanotubes
\cite{Jorio1}.

For a given tube, we first relax the atomic positions and allow the
lattice constant to adjust to minimize the energy. We then calculate
the RBM energy. The results for all tubes in the diameter range from
0.5 to 4.0 nm are shown in Fig.~\ref{fig-rbm}. Simple scaling
arguments based on continuum elasticity suggest that RBM energies
should scale with diameter as $\hbar\omega_{RBM}=A/d$. Experimental
data are typically fitted with the phenomenologically adapted form
$\hbar\omega_{RBM}=A/d+B$. For tubes of $d$=1 nm, experimental
phonon energies A+B are reported in the range 226-248 cm$^{-1}$
\cite{rbm0,rbm10,rbm2,rbm3,rbm6,rbm9,rbm11}, while {\it ab-initio}
calculations suggest $A+B$=234 or 226 cm$^{-1}$ \cite{rbm0,rbm10}.
The constant off-set was reported in the range from $B=-6$ cm$^{-1}$
to $B=27$ cm$^{-1}$. Recently, it was reported \cite{rbm11} that a
non-zero offset $B$ is caused by the interaction with a substrate,
while for freely suspended nanotubes $B$ should be zero.

Within our model, the RBM mode shows accurate $A/d$ scaling
independent of chirality, with $B \approx 0$ and $A\approx 225$ cm$^{-1}$
(where $d$ is in nm) as shown in Fig.~\ref{fig-rbm}.
From the theory of elasticity, $A=2\hbar v_L$, which gives
$A\approx 225$ cm$^{-1}$ for the parameters of Table~\ref{tab1},
in accord with the numerical result.  The model is in good
agreement with the most recent experimental \cite{rbm11} and
theoretical  \cite{rbm10} values of $A=$227 and $A+B=$226 cm$^{-1}$
respectively.

In general, a valence force model will have some anharmonicity.
Since we have not attempted to fit experimental or \textit{ab initio}
anharmonicities, any anharmonicities are likely to be unphysical.
It is therefore desirable to minimize the anharmonicity in the model,
and the form of Eq.~(\ref{energy}) is designed with this in mind.
One measure of anharmonicity is the Gruneisen parameter
$\gamma=-  ( 2\omega)^{-1}(d\omega / d\varepsilon)$,
which represents the fractional shift in phonon frequency $\omega$
when the crystal is subjected to a strain $\varepsilon$ in all
directions. For the doubly degenerate $E_{2g}$ phonon mode in
graphene, our model gives $\gamma_{E_{2g}}\approx -0.2$. This is
much smaller in magnitude than the experimental value of
$\gamma_{E_{2g}}\approx 2.0$ \cite{Rubio,Ferrari2}, confirming that
our model is relatively harmonic in this respect.

For nanotubes, we have another form of anharmonicity, the phonon
shifts due to bending of the graphene sheet. We have calculated the
shifts in LO and TO phonons relative to graphene. The TO mode shift
is less than 12 cm$^{-1}$/d$^2$ in our model (where $d$ is in nm),
and the LO mode and the LO-TO splitting are even less. Experimental
shifts are four times larger in semiconducting nanotubes
\cite{JorioLOTO}, confirming that our model successfully minimizes
any unintended anharmonicities.

In conclusion, we have developed a valence force model applicable
for $sp^2$-carbon based structures. Our model gives a good fit of
the graphene phonon dispersion and elastic constants, and describes
well the RBM energy of nanotubes. The model also avoids the
unphysical strong anharmonicities that occur in some valence force
models. Most importantly, in contrast to other phonon models for
$sp^2$-bonded carbon, Eq.~(\protect{\ref{energy}}) makes no
reference to an underlying lattice, so it can be directly applied to
distorted geometries.

We gratefully acknowledge N.~Marzari, O.~Dubay, and D.~Kresse for
providing data for Figs.~\ref{fig-fit} and \ref{fig-fit1},
and W.~A.~Harrison and A.~Jorio for helpful discussions.


\begin{thebibliography}{9}
\bibitem{footnote1} The model parameters $\beta_{r1}=19.39$, $\beta{c}=3.409$,
$\beta_{v}=1.318$, $\beta_{r2}=8.364$, $\beta_{p}=0.008078$, and
$\beta_{rc}=2.534$ (in eV) give the fit to LDA calculations shown in
Fig.\ref{fig-fit1}, with the standard deviation of $\sigma=22.6$
cm$^{-1}$ and the maximum deviation of $\vert
\hbar\omega_{model}-\hbar\omega_{LDA}\vert=57$ cm$^{-1}$. A similar
fit quality can be achieved in fitting GGA phonon dispersions from
Ref.~\protect{\cite{Marzari}}.  However, we recommend using instead
the parameters of Table~\ref{tab1}.
\end{thebibliography}
\end{document}